# Mesons with a light quark-antiquark pair and the Bethe-Salpeter equation

Dean Lee[1]

Physics Department, Harvard University, Cambridge MA 02138

**Abstract**

The light quark-antiquark scattering Green's function is considered near a meson resonance peak. The Bethe-Salpeter equation is used to write formal expressions for the resonance width/mass ratio. Arguments are made concerning to what extent this ratio can be calculated perturbatively, and an upper bound is placed on the growth of this ratio as a function of radial excitation. Certain mesons and their radial excitations are considered, as well as the more general issue of classifying mesons in the quark model.

## 1. Introduction

One interesting feature of QCD is the existence of a large number of meson and baryon resonances. Unfortunately confinement, an essential characteristic of bound states in QCD, is not completely understood and most quantitative information about these bound state resonances is deduced from symmetry arguments such as SU(3) flavor symmetry, chiral symmetry, and heavy quark symmetry, or from empirical models such as quark-antiquark potential models or flux-tube models.

The following analysis is an attempt to deduce some consequences of perturbative QCD for light quark-antiquark meson resonances. The Bethe-Salpeter equation, written to lowest order in the perturbative expansion, is used to extract information about quark-antiquark scattering near resonance peaks. Perturbative approximations to the full Bethe-Salpeter equation miss important confinement effects, and a straightforward perturbative approximation is avoided. Instead, perturbation theory is employed on the expectation values of operators interposed between exact wavefunctions. The assumption is made, and motivated later, that the expectation values of non-perturbative operators associated with confinement can be neglected.

## 2. Quark-Antiquark Scattering and the Bethe-Salpeter Equation

The Green's function of interest is

$$G_{2\bar{1}}^{\alpha\beta,\mu\nu}(p;-q',q)\cdot\delta^4(p-p')=\left\langle 0\left|\,\mathrm{T}\left\{\overline{\psi}_{1i}^{\alpha}\left(\frac{-p'-q'}{2}\right)\psi_{2}^{\beta i}\left(\frac{-p'+q'}{2}\right)\overline{\psi}_{2j}^{\mu}\left(\frac{p-q}{2}\right)\psi_{1}^{\nu j}\left(\frac{p+q}{2}\right)\right\}\right|0\right\rangle_{\text{connected}} \tag{2.1}$$

[1]Supported by the Hertz Foundation and the National Science Foundation under Grant #PHY9218167

where the Greek superscripts are Dirac indices, the contracted Latin indices show the color singlets, and the subscripts 1, 2 refer to quark flavor. Only connected diagrams contributing to this process will be considered, as they should be sufficient to describe properties of the meson resonance peak.

As this Green's function is gauge dependent, many of the quantities derived from it will also be gauge dependent, including the Bethe-Salpeter wavefunctions. However, physical quantities such as masses and widths will, as usual, be gauge independent. Unfortunately, the rather useful process of checking the gauge invariance of physical quantities explicitly will not be possible, because wavefunctions are not calculated.

Two particle irreducible (2PI) diagrams are defined as those that do not contain a purely quark-antiquark intermediate state. By convention, the external quark and antiquark lines of 2PI diagrams are removed. Below is a diagrammatic recurrence relation, known as the Bethe-Salpeter equation, for the full connected Green's function written in terms of 2PI diagrams and full quark and antiquark propagators [1].

$$(I)$$

By using a matrix shorthand in which matrix multiplication signifies contracting Dirac indices and integrating over momenta, the recurrence relation can be written succinctly and solved.

$$G = K + KIG \qquad (2.2)$$

$$G = \left( K^{-1} - I \right)^{-1} \qquad (2.3)$$

$G \equiv$ full connected Green's function
$K \equiv$ full quark and antiquark propagators
$I \equiv$ two particle irreducible Green's function

Let $\phi_i$ be the eigenfunctions of $G$, and $\lambda_i$ be the respective eigenvalues.

$$G\phi_i = \lambda_i \phi_i \qquad (2.4)$$

It should be noted that the matrix $G$ will, in general, not be hermitian. Let $\tilde{\phi}_i$, represented as a row vector, be the eigenfunctions of $G^T$, and $\tilde{\lambda}_i$ be the respective eigenvalues.

$$\tilde{\phi}_i G = \tilde{\lambda}_i \tilde{\phi}_i \qquad (2.5)$$

As any matrix and its transpose share the same characteristic equation, they share identical eigenvalues. Let $M$ be the subspace spanned by the nondegenerate eigenfunctions of $G$. In this subspace, the indexing of both sets of eigenfunctions can thus be aligned by matching eigenvalues.

$$\text{index } i: \quad G\phi_i = \lambda_i \phi_i, \quad \tilde{\phi}_i G = \lambda_i \tilde{\phi}_i \tag{2.6}$$

$$\text{Since } \lambda_i \tilde{\phi}_i \phi_k = \tilde{\phi}_i G \phi_k = \lambda_k \tilde{\phi}_i \phi_k \Rightarrow \tilde{\phi}_i \phi_k = c_i \delta_{ik}. \tag{2.7}$$

Let $S(\tilde{S})$ be the subspace of $M$ spanned by $\phi_i(\tilde{\phi}_i)$ for $c_i$ nonzero. Using orthogonality relations, the eigenfunctions can be used to write a bilinear form, $G'$, equivalent to $G$ on these restricted subspaces.

$$G' \equiv \sum_i \lambda_i \left( \tilde{\phi}_i \phi_i \right)^{-1} \phi_i \tilde{\phi}_i \tag{2.8}$$

$$\tilde{f} G g = \tilde{f} G' g \quad \forall \tilde{f} \in \tilde{S}, \quad g \in S \tag{2.9}$$

By the Bethe-Salpeter relation derived above,

$$\lambda_i^{-1} = \left( \tilde{\phi}_i \phi_i \right)^{-1} \tilde{\phi}_i G^{-1} \phi_i = \left( \tilde{\phi}_i \phi_i \right)^{-1} \tilde{\phi}_i \left( K^{-1} - I \right) \phi_i = \left( \tilde{\phi}_i \phi_i \right)^{-1} \left( \tilde{\phi}_i K^{-1} \phi_i - \tilde{\phi}_i I \phi_i \right) \tag{2.10}$$

$$G' \equiv \sum_i \left( \tilde{\phi}_i K^{-1} \phi_i - \tilde{\phi}_i I \phi_i \right)^{-1} \phi_i \tilde{\phi}_i \tag{2.11}$$

## 3.  The Resonance Peak

By momentum conservation, the eigenfunctions described above can be chosen to be states of definite total momentum. To avoid confusion, this total momentum will be written as a subscript to make explicit its role as an index and to distinguish it from the relative momentum in whose Hilbert space the eigenfunctions reside.

$$G'(p; -q', q) = \sum_i \frac{\phi_{i,-p}(-q') \tilde{\phi}_{i,p}(q)}{f_i(p^2)} \tag{3.1}$$

$$\text{where} \quad f_i(p^2) \equiv \iint d^4q \, d^4q' \left( \tilde{\phi}_{i,p}(q') K^{-1}(p; -q', q) \phi_{i,-p}(-q) - \tilde{\phi}_{i,p}(q') I(p; -q', q) \phi_{i,-p}(-q) \right) \tag{3.2}$$

Assuming that $f_i(p^2)$ is well-defined and that $\left| f_i(p^2) \right|$ has a unique minimum, the location of that minimum will be denoted

$$p_{i,\min}^2: \quad \left| f_i(p_{i,\min}^2) \right| = \min_{p^2} \left\{ \left| f_i(p^2) \right| \right\} \tag{3.3}$$

If $\left| f_i(p_{i,\min}^2) \right|$ is small then $G'(p; -q', q)$ near $p = p_{i,\min}^2$ is dominated by

$$\left[ f_i(p^2) \right]^{-1} \phi_{i,-p}(-q') \tilde{\phi}_{i,p}(q) \tag{3.4}$$

By adjusting the eigenfunction's phase, the behavior of $f_i(p^2)$ near $p_{i,\min}^2$ can be written as

$$f_i(p^2) \approx a \left( p^2 - p_{i,\min}^2 \right) + i \, b \, p_{i,\min}^2 \quad a, b \text{ real} \tag{3.5}$$

The interpretation of these results is that the square of the mass of the resonance is $p_{i,\min}^2$. The width/mass ratio of the resonance is $b/a$.

## 4. Perturbative Calculations

This section applies the formalism of the preceding sections to meson resonances. Several assumptions and approximations are made. These assumptions and approximations are motivated by qualitative physical arguments, but the actual size of the errors incurred are not known.

Even for highly excited meson resonances, where energies are sufficient to suggest that the quark-antiquark interaction is predominantly in the asymptotically free regime, confinement must still modify the behavior of the wavefunction in the region where the quark and antiquark are widely separated. The present analysis adopts the view that for such highly excited meson resonances, confinement can be effectively modeled as a large distance boundary condition on the wavefunction.

If this confinement boundary condition is sufficiently abrupt and severe, a high wall for instance, the wavefunction will not penetrate substantially into regions in which confinement is important. A simple analogy can be made using non-relativistic quantum mechanics in one dimension. If one starts with a smooth potential $V_0$ and adds to it a pair of infinite walls at locations $a$ and $-a$, the eigenstates of the new potential, $V$, will vanish at $a$ and $-a$. For the purposes of calculating the expectation value of $V$, the walls can be neglected, as, in some sense, information about the walls are contained in the wavefunctions themselves. It will be assumed that something similar happens in the case of highly excited mesons, and the expectation values of non-perturbative, confinement-producing operators are neglected.

All perturbative calculations are done in Landau gauge and dimensionally regulated in the $\overline{MS}$ prescription. Light quark masses are neglected although, to some extent, they make a reprise in phase space corrections described later. The running strong coupling constant, renormalized at the mass of the resonance under consideration, $m_I$, is used to minimize higher loop corrections. In general, diagrams are kept to lowest order. Free quark and antiquark propagators are used in the matrix $K$. A freely propagating t-channel gluon is the lowest order contribution to the dispersive part of the matrix $I$. The absorptive part of the matrix $I$, however, is a more subtle matter and gets its lowest order contribution from a higher order diagram.

The absorptive part of any process is associated with intermediate states situated on their physical mass shells. As the physical states of the underlying free theory are quite different from the physical states of fully-interacting QCD, naive lowest order perturbative calculations of the absorptive part of a process can be misleading. For instance, in calculating the propagator for meson resonances, the lowest lying possible physical intermediate states in the fully-interacting theory are two meson states. In order for the intermediate two meson state to be produced, an extra quark-antiquark pair must first be produced since a two meson state has two valence quark-antiquark pairs. The simplest approach then, and the one followed here, is that the lowest order contribution to the absorptive part of a strongly interacting process is given by the lowest order diagram(s) with the correct valence particle structure. This will be referred to as valence correspondence. In general, three body or higher body decays occur less frequently, and only decays into two mesons are treated in this analysis. . Below are the lowest order diagrams and calculations for the expectation values of equation (3.2).

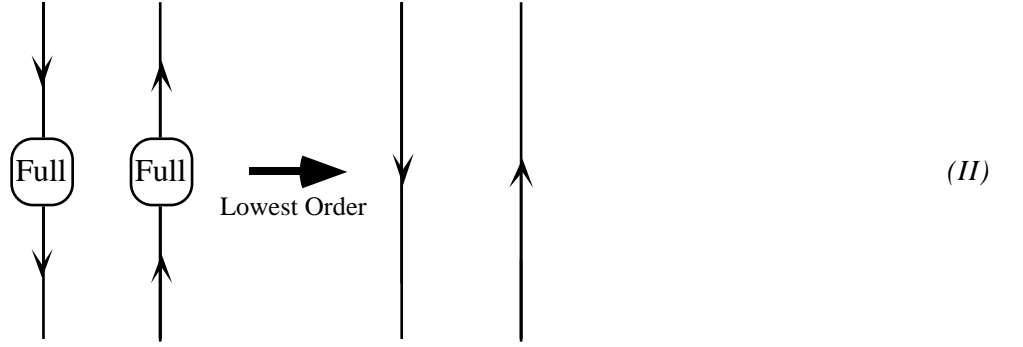

*(II)*

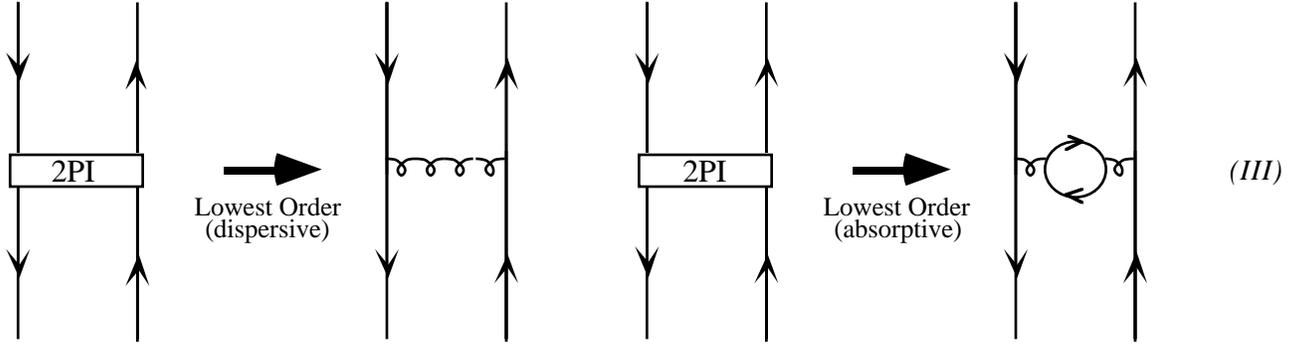

*(III)*

Evaluating these diagrams,

$$\kappa_i(p^2) \equiv \tilde{\phi}_{i,p} K^{-1} \phi_{i,-p} \approx \tfrac{1}{2} \cdot \int d^4q \left( -\text{tr}\left( \tilde{\phi}_{i,p}(q) [\not{p} - q] \phi_{i,-p}(-q) [\not{p} + q] \right) \right) \quad \text{(phase chosen to be real)} \tag{4.1}$$

$$I_i(p^2) \equiv \tilde{\phi}_{i,p} I \phi_{i,-p} \approx 4 \cdot \iint d^4q \, d^4q' \left\{ \text{Re}\left[ g^2(m_i^2) \cdot M_p(q,q') \right] + i \cdot \frac{n_f}{24\pi^2} g^4(m_i^2) \cdot \text{Im}\left( \ln\left( \frac{-(q-q')^2 - i\varepsilon}{m_i^2} \right) \right) \cdot M_p(q,q') \right\} \tag{4.2}$$

$$\text{where } M_p(q,q') \equiv i \cdot \left( \frac{g^{\mu\nu}}{(q-q')^2 + i\varepsilon} - \frac{(q-q')^\mu (q-q')^\nu}{\left( (q-q')^2 + i\varepsilon \right)^2} \right) \cdot \text{tr}\left[ \tilde{\phi}_{i,p}(q') \gamma_\mu \phi_{i,-p}(-q) \gamma_\nu \right] \tag{4.3}$$

At this point, it is convenient to define some auxiliary functions,

$$\chi_i(p^2) \equiv \text{Re}\left\{ \tilde{\phi}_{i,p} I \phi_{i,-p} \right\} = \chi_i^+(p^2) + \chi_i^-(p^2) \tag{4.4}$$

$$\pi_i(p^2) \equiv \text{Im}\left\{ \tilde{\phi}_{i,p} I \phi_{i,-p} \right\} = i \cdot n_f \cdot \frac{g^2(m_i^2)}{24\pi} \chi_i^+(p^2) \tag{4.5}$$

$$\text{where } \chi_i^\pm(p^2) \equiv 4 \cdot \text{Re}\left\{ \iint d^4q \, d^4q' \left[ g^2(m_i^2) \cdot \theta\left( \frac{\pm(q-q')^2}{m_i^2} \right) \cdot M_p(q,q') \right] \right\} \tag{4.6}$$

Equation (3.2) can now be written as

$$f_i(p^2) = \kappa_i(p^2) - \left[ \chi_i(p^2) = \chi_i^+(p^2) + \chi_i^-(p^2) \right] + \left[ \pi_i(p^2) = i \cdot n_f \cdot \frac{g^2(m_i^2)}{24\pi} \chi_i^+(p^2) \right] \tag{4.7}$$

$n_f$ refers to the number of possible flavors of light quark loops in the absorptive diagram shown in figure *(III)*. The simplest approach is to set $n_f = 3$. It may be useful, however, to be more careful. Although the masses of light quarks have been neglected in the above calculations, the mass of the light quark does become important near a kinematic threshold. One example is the $\rho(770)$, which can decay into two pions but is kinematically forbidden to decay into two kaons. To satisfy valence correspondence (VC), the $s$-quark loop should not be included in counting $n_f$ for the $\rho(770)$, since there are no kinematically allowed two meson states with the correct valence properties. In order to handle such kinematic effects systematically, the following formula, motivated by the phase space factors for two body decays, will be used to evaluate $n_f$.

$$n_f \approx \sum_{\text{flavor of } q\bar{q}} \max_{\substack{\text{two meson states} \\ (\text{masses } m_1, m_2) \text{ obeying VC}}} \left( \frac{\sqrt{\left(m_i^2 - (m_1 + m_2)^2\right)\left(m_i^2 - (m_1 - m_2)^2\right)}}{m_i^2} \right) \tag{4.8}$$

It is useful to further analyze $\kappa_i(p^2)$. By charge conjugation and SU(3) flavor symmetry, terms linear in $q$ in equation (4.1) do not contribute.

$$\kappa_i(p^2) \approx \kappa_i^P(p^2) + \kappa_i^Q(p^2) \tag{4.9}$$

$$\text{where} \quad \kappa_i^P(p^2) \equiv \tfrac{1}{4} \cdot \int d^4q \; \text{tr}\left[ -\tilde{\phi}_{i,p}(q)\, \not{p}\, \phi_{i,-p}(-q)\, \not{p} \right] \tag{4.10}$$

$$\text{and} \quad \kappa_i^Q(p^2) \equiv \tfrac{1}{4} \cdot \int d^4q \; \text{tr}\left[ -\tilde{\phi}_{i,p}(q)\, \not{q}\, \phi_{i,-p}(-q)\, \not{q} \right] \tag{4.11}$$

The total spin of quark-antiquark mesons arise from the intrinsic spin of the quark and antiquark as well as their orbital angular momentum. It is known that the intrinsic spin of the quark-antiquark pair is either spin one or zero. This fact is used to simplify equation (4.10).

$$\phi_{i,-p} = \varphi_{i,-p}^P \cdot \gamma_5 + \varphi_{i,-p}^{V\mu} \cdot \gamma_\mu + \varphi_{i,-p}^S \cdot 1 + \varphi_{i,-p}^{A\mu} \cdot \gamma_\mu \cdot \gamma_5 \tag{4.12}$$

$$\tilde{\phi}_{i,-p} = \tilde{\varphi}_{i,-p}^P \cdot \gamma_5 + \tilde{\varphi}_{i,-p}^{V\mu} \cdot \gamma_\mu - \tilde{\varphi}_{i,-p}^S \cdot 1 + \tilde{\varphi}_{i,-p}^{A\mu} \cdot \gamma_\mu \cdot \gamma_5 \tag{4.13}$$

$$\kappa_i^P(p^2) \equiv p^2 \cdot d_i(p^2) \tag{4.14}$$

$$\text{where} \quad d_i(p^2) \equiv \int d^4q \left[ \tilde{\varphi}_i^P(q)\varphi_i^P(-q) + \tilde{\varphi}_i^S(q)\varphi_i^S(-q) + \tilde{\varphi}_i^{V\mu}(q)\varphi_{i\mu}^V(-q) + \tilde{\varphi}_i^{A\mu}(q)\varphi_{i\mu}^A(-q) \right] \tag{4.15}$$

The next step is to write down an expression for $f_i(p^2)$ near $p^2 = m_i^2$. Using the orthogonality condition derived in equation (2.7) and assuming completeness of the Bethe-Salpeter eigenfunctions at $p^2 = m_i^2$, $f_i(p^2)$ can be expanded in terms of eigenfunctions at $p^2 = m_i^2$. The result, to first order in $p^2 - m_i^2$, is simply

$$f_i(p^2) \approx \iint d^4q \; d^4q' \left( \tilde{\phi}(q') K^{-1}(p; -q', q)\phi(-q) - \tilde{\phi}(q') I(p; -q', q)\phi(-q) \right) \tag{4.16}$$

$$\text{where} \quad \phi_i \equiv \phi_{i, p^2 = m_i^2} \quad \text{and} \quad \tilde{\phi}_i \equiv \tilde{\phi}_{i, p^2 = m_i^2} \tag{4.17}$$

This can be rewritten in terms of the various auxiliary functions defined above,

$$f_i(p^2) \approx \frac{p^2}{m_i^2} \cdot \kappa_i^P(m_i^2) + \kappa_i^Q(m_i^2) - \chi_i^+(m_i^2) - \chi_i^-(m_i^2) + \pi(m_i^2) \tag{4.18}$$

$$\kappa_i^P(m_i^2) + \kappa_i^Q(m_i^2) - \chi_i^+(m_i^2) - \chi_i^-(m_i^2) \approx 0 \tag{4.19}$$

The resonance mass/width ratio, $r_i$, is given by

$$r_i = \frac{\frac{1}{2} \cdot \pi\left(m_i^2\right)}{\kappa_i^P\left(m_i^2\right)} \tag{4.20}$$

Using equations (4.8), (4.14), and (4.19),

$$r_i \approx \frac{\frac{g^2\left(m_i^2\right)}{24\pi}\chi_i\left(m_i^2\right)}{-\kappa_i^Q\left(m_i^2\right)+\chi_i^+\left(m_i^2\right)+\chi_i^-\left(m_i^2\right)} \cdot \sum_{\text{flavor of } q\bar{q}} \max_{\substack{\text{two meson states} \\ \text{(masses } m_1, m_2) \text{ obeying VC}}}\left(\frac{\sqrt{\left(m_i^2-\left(m_1+m_2\right)^2\right)\left(m_i^2-\left(m_1-m_2\right)^2\right)}}{m_i^2}\right) \tag{4.21}$$

$$m_i^2 = \frac{\kappa_i^P\left(m_i^2\right)}{d_i\left(m_i^2\right)} = \frac{-\kappa_i^Q\left(m_i^2\right)+\chi_i^+\left(m_i^2\right)+\chi_i^-\left(m_i^2\right)}{d_i\left(m_i^2\right)} \tag{4.22}$$

## 5. Radial Excitations

It is convenient to define the normalized quantities,

$$\hat{\kappa}_i^Q\left(m_i^2\right) \equiv \frac{\kappa_i^Q\left(m_i^2\right)}{d_i\left(m_i^2\right)} \tag{5.1}$$

$$\hat{\chi}_i^{\pm}\left(m_i^2\right) \equiv \frac{\chi_i^{\pm}\left(m_i^2\right)}{d_i\left(m_i^2\right)} \tag{5.2}$$

The remainder of this analysis will be concerned with the behavior of these three quantities as a function of radial excitation number, $n$, with all other parameters fixed. It is assumed that the mass of any sequence of radial excitations of resonances is, for sufficiently large $n$, well approximated by a smooth, monotonically increasing function of $n$. Since $-\hat{\kappa}_i^Q\left(m_i^2\right)+\hat{\chi}_i^+\left(m_i^2\right)+\hat{\chi}_i^-\left(m_i^2\right) = m_i^2$, it is also assumed that each of $-\hat{\kappa}_i^Q\left(m_i^2\right)$, $\hat{\chi}_i^+\left(m_i^2\right)$, and $\hat{\chi}_i^-\left(m_i^2\right)$ can be approximated by smooth functions of $n$. Furthermore, it is naively assumed that for large $n$ these functions are well-behaved and do not oscillate rapidly nor with wide amplitudes (e.g., power law or logarithmic).

The following definition will be useful,

$$k_i \equiv \frac{r_i}{\frac{g^2\left(m_i^2\right)}{24\pi} \cdot \sum_{\text{flavor of } q\bar{q}} \max_{\substack{\text{two meson states} \\ \text{(masses } m_1, m_2) \text{ obeying VC}}}\left(\frac{\sqrt{\left(m_i^2-\left(m_1+m_2\right)^2\right)\left(m_i^2-\left(m_1-m_2\right)^2\right)}}{m_i^2}\right)} \tag{5.3}$$

$$\approx \frac{\chi_i^-\left(m_i^2\right)}{-\kappa_i^Q\left(m_i^2\right)+\chi_i^+\left(m_i^2\right)+\chi_i^-\left(m_i^2\right)} \tag{5.4}$$

As $n$ becomes large, one possibility is that $\hat{\chi}_i^-\left(m_i^2\right)$ dominates $-\hat{\kappa}_i^Q\left(m_i^2\right)$ and $\hat{\chi}_i^+\left(m_i^2\right)$. Sequences of radial excitations of meson resonances with this property, if there are any, will be denoted as type 1. Type 1 sequences have the property that for sufficiently large $n$, $k_i \rightarrow 1$.

A second possibility is that $\hat{\chi}_i^-\left(m_i^2\right)$ is dominated by either $-\hat{\kappa}_i^Q\left(m_i^2\right)$ or $\hat{\chi}_i^+\left(m_i^2\right)$. These resonance sequences will be denoted type 2 and have the property that for sufficiently large $n$, $k_i \rightarrow 0$.

The third possibility is that $\hat{\chi}_i^2(m_i^2)$ shares dominance with one or more of the other two terms, i.e., has the same large $n$ behavior. Resonance sequences with this property will be denoted type 3. For large $n$, type 3 sequences will have $k_i \to k$, not necessarily 0 or 1.

For all three types of sequences, $k_i$ is bounded above. This is simply the statement that the denominator in equation (5.4) is not fine-tuned to be arbitrarily smaller than the numerator. The bound on $k_i$ gives a bound on the growth of $r_i$.

$$\exists k > 0 \quad s.t. \quad r_i \le k \cdot \frac{g^2(m_i^2)}{24\pi} \cdot \sum_{\substack{\text{flavor of } q\bar{q}}} \max_{\substack{\text{two meson states} \\ (\text{masses } m_1, m_2) \text{ obeying VC}}} \left( \frac{\sqrt{\left(m_i^2 - (m_1 + m_2)^2\right)\left(m_i^2 - (m_1 - m_2)^2\right)}}{m_i^2} \right) \tag{5.5}$$

## 6. Experimental Data

There are a number of mesons for which a reasonable quark model interpretation exists, and the initial entries of some radial excitation sequences can be examined. The data for these is presented below [2]. In the calculation of $k_i$, $\Lambda_{\overline{MS}}$ is taken to be 250 MeV.

| Meson | $I^G(J^{PC})$ | $n\ ^{2S+1}L_J$(quark model) | $r_i$ | $k_i$ |
|---|---|---|---|---|
| $\rho(770)$ | $1^+(1^-)$ | $1\ ^3S_1$ | $.197 \pm .0016$ | $1.02 \pm .01$ |
| $\rho(1450)$ | $1^+(1^-)$ | $2\ ^3S_1$ | $.212 \pm .041$ | $1.19 \pm .23$ |
| $K^*(892)$ | $\frac{1}{2}(1^-)$ | $1\ ^3S_1$ | $.0558 \pm .0009$ | $.471 \pm .008$ |
| $K^*(1410)$ | $\frac{1}{2}(1^-)$ | $2\ ^3S_1$ | $.161 \pm .016$ | $.996 \pm .099$ |
| $\phi(1020)$ | $0^-(1^-)$ | $1\ ^3S_1$ | $.00434 \pm .00006$ | $.103 \pm .001$ |
| $\phi(1680)$ | $0^-(1^-)$ | $2\ ^3S_1$ | $.0893 \pm .0298$ | $.616 \pm .206$ |
| $\omega(782)$ | $0^-(1^-)$ | $1\ ^3S_1$ | $.0108 \pm .0001$ | (?) no 2-body decays |
| $\omega(1390)$ | $0^-(1^-)$ | $2\ ^3S_1$ | $.123 \pm .042$ | $.880 \pm .304$ |
| $f_2(1270)$ | $0^+(2^+)$ | $1\ ^3P_2$ | $.146 \pm .016$ | $.793 \pm .087$ |
| $f_2(1810)$ | $0^+(2^+)$ | $2\ ^3P_2$ | $.214 \pm .010$ | $1.30 \pm .06$ |
| $f_2'(1525)$ | $0^+(2^+)$ | $1\ ^3P_2$ | $.0498 \pm .0066$ | $.350 \pm .046$ |
| $f_2'(2010)$ | $0^+(2^+)$ | $2\ ^3P_2$ | $.100 \pm .032$ | $.697 \pm .32$ |
| $K_2^*(1430)$ | $\frac{1}{2}(2^+)$ | $1\ ^3P_2$ | $.0688 \pm .0016$ | $.427 \pm .010$ |
| $K_2^*(1980)$ | $\frac{1}{2}(2^+)$ | $2\ ^3P_2$ | $.197 \pm .020$ | $1.29 \pm .13$ |
| $\pi$ | $1^-(0^+)$ | $1\ ^1S_0$ | stable | $\infty$ in chiral limit |
| $\pi(1300)$ | $1^-(0^+)$ | $2\ ^1S_0$ | $.338 \pm .062$ | $3.81 \pm .70$ |
| $\pi(1770)$ | $1^-(0^+)$ | $3\ ^1S_0$ | $.175 \pm .028$ | $1.34 \pm .21$ |
| $K$ | $\frac{1}{2}(0^-)$ | $1\ ^1S_0$ | stable | $\infty$ in chiral limit |
| $K(1460)$ | $\frac{1}{2}(0^-)$ | $2\ ^1S_0$ | $\sim .178$ | $\sim 1.96$ |
| $K(1830)$ | $\frac{1}{2}(0^-)$ | $3\ ^1S_0$ | $\sim .137$ | $\sim 1.11$ |

The chiral limit divergence of $k_i$ for $\pi$ and $K$ is obtained by noticing that the denominator in equation (5.8) vanishes as $m_\pi^2$ or $m_K^2$, whereas the numerator, in general, need not vanish. It may be argued, qualitatively, that the large values of $k_i$ for the $2\,^1S_0$ pseudoscalars is a consequence of the relative smoothness of $k_i$ as a function of radial excitation.

Higher excitations of mesons will have to be examined in order to check if $k_i$ does actually appear to converge and, if so, what $k_i$ converges to in order to classify the excitation sequence as type 1, 2, or 3.

## 7.  Remarks

The main theoretical result of this analysis is that the width of mesons in a radial excitation sequence does not grow arbitrarily large with respect to meson mass, and the upper bound on the growth of the width is given in equation (5.5). It is worth noting that this result should be valid, with slight changes of parameters, for confining QCD-like theories such as technicolor.

If in the future a radial excitation sequence of reasonable length is identified, classifying this sequence as type 1, 2, or 3 may be useful in providing some information about the distribution of the wavefunction in momentum space. For immediate purposes, however, it may prove useful to employ the converse and assume that $k_i$ is convergent in order to check whether a given meson is a good candidate for the next radial excitation of a given sequence. In this regard, one might have a potentially useful tool to help classify mesons in the quark model. Unlike SU(3) flavor or nonet symmetry which allows "horizontal" comparisons of mesons within an approximately degenerate multiplet, this technique would be a "vertical" comparison of radial excitations of mesons with sequentially increasing mass.

## Acknowledgment

I am grateful to my advisor, Howard Georgi, for his generous guidance, patience, and encouragement.

## References

1.  Salpeter, E. E. and Bethe, H. A., Phys. Rev. 84 (1951), 1232.
2.  Particle Data Group, Review of Particle Properties, Phys. Rev. D 50 (1994).